\documentclass[aps,two column]{revtex4}
\usepackage{amsmath}
\usepackage{color}
\usepackage{psfrag}
\usepackage{dcolumn}
\usepackage{bm}
\usepackage[latin1]{inputenc}
\usepackage[spanish,english]{babel}
\usepackage{amsfonts}
\usepackage{amssymb,epsf}
\usepackage{graphicx}
\usepackage{epstopdf}

\begin{document}

\title{Three-dimensional energy-dependent $C$-metric: \\
black hole solutions}
\author{B. Eslam Panah$^{1,2,3}$\footnote{
email address: eslampanah@umz.ac.ir}}
\affiliation{$^{1}$ Sciences Faculty, Department of Physics, University of Mazandaran, P.
O. Box 47415-416, Babolsar, Iran\\
$^{2}$ ICRANet-Mazandaran, University of Mazandaran, P. O. Box 47415-416,
Babolsar, Iran\\
$^{3}$ ICRANet, Piazza della Repubblica 10, I-65122 Pescara, Italy}

\begin{abstract}
Considering a three-dimensional $C$-metric and adding energy-dependent to
this spacetime, we first create a three-dimensional energy-dependent $C$%
-metric. Then, we extract accelerating BTZ black hole solutions in gravity's
rainbow. Besides, we show that (A)dS black holes cover by an event horizon
that depends on all the parameters of this theory. Using the definition of
Hawking temperature, we obtain the temperature of these black holes and
study the effects of various parameters on this quantity. We find a critical
radius in which the temperature is always positive (negative) before (after)
it. Then, we obtain the entropy of such black holes. Our analysis indicates
that there is the same behavior for entropy, similar to the temperature.
Indeed, before (after) the critical radius, the entropy is positive
(negative). In order to study the local stability of such black holes, we
calculate the heat capacity. We find two different behaviors for the heat
capacity, which depend on the cosmological energy-dependent constant. As a
final result, accelerating AdS BTZ black holes can satisfy the physical
condition and local stability at the same time.
\end{abstract}

\maketitle

%%%%%%%%%%%%%%%%%%%%%%%%%%%%%%%%%%%%%%%%%%%%%%%%%%%%%%%%%%%%%%%%%%%%%%%%%%%%%%%%%%%%%%%%%%%%%%%%%%%%%%%%%%%

%%%%%%%%%%%%%%%%%%%%%%%%%%%%%%%%%%%%%%%%%%%%%%%%%%%%%%%%%%%%%%%%%%%%%%%%%%%%%%%%%%%%%%%%%%%%%%%%%%%%%%%%%%%

\section{\noindent Introduction}

Three-dimensional gravity has attracted much attention since the late 20th
century, especially after works by Deser, Jackiw, and 't Hooft \cite%
{DJHI,DJHII,DJHIII}, and Witten \cite{WittenI}. Especially in quantum
gravity, three-dimensional theories of gravity can be a useful tool. In
other words, we can use it to examine the classical and quantum dynamics of
point sources. Also, the probability of the existence of a connection
between three-dimensional gravity and the Chern-Simons theory has been
studied \cite{Achucarro1986,WittenII}. In this regard, Banados, Teitelboim,
and Zanelli (BTZ) extracted a black hole solution in three-dimensional
spacetime in general relativity (GR) with the negative cosmological
constant, which is known as BTZ black hole \cite{BTZ}. The three-dimensional
black hole has thermodynamic properties much like those of a
four-dimensional black hole. In this regard, Witten obtained the entropy of
the BTZ black holes \cite{WittenIII}. In addition, the study of
three-dimensional black holes can help our understanding, for example, a
profound insight into the physics of black holes, the quantum view of
gravity, its relations to string theory, AdS/CFT correspondence, loss of
information, and the endpoint of quantum evaporation \cite%
{Import1,Import2,Import3,AdS1,AdS2,Harvey}. Based on this, many works have
been done on BTZ black holes which here I mention some of them \cite%
{BTZ1,BTZ2,BTZ3,BTZ4,BTZ6,BTZ7,BTZ8,BTZ9,BTZ10b,BTZ11,BTZ12,BTZ13,BTZ14,BTZ15}%
.

On the other hand, the black hole solutions which are extracted in the $C$%
-metric \cite{Kinnersley,Plebanski,Dias,Griffiths} are known as accelerating
black holes. Indeed, by considering the $C$-metric, the accelerating black
holes are obtained in the solution to Einstein's equation with negative
cosmological constant \cite{Plebanski}. These black holes have different
behaviors from other famous black hole solutions, which are related to the
properties of the $C$-metric. One of these different behaviors is related to
the existence of a conical deficit angle along one polar axis that provides
the force driving the acceleration. Besides, according to the AdS/CFT
correspondence, the $C$-metric has fruitful applications, for example, the
construction of exact black holes on the branes \cite{Emparan2000}, the
black funnels and droplets \cite{Hubeny2010}, the plasma ball \cite%
{Emparan2009}, the quantum BTZ black hole \cite{Emparan2020}, and the
spinning spindles from accelerating black holes \cite{Ferrero2021}. In 
addition, it was indicated that accelerated black holes have a higher
Hawking temperature than the Unruh temperature of the accelerated frame \cite%
{Letelier}. Also, their asymptotic behaviors are very complicated and
dependent on their various parameters. In a valuable work, the first
law of black hole thermodynamics from the charged accelerating black holes
with the usual identification of entropy proportional to the area of the
event horizon was obtained in Ref. \cite{The1}. Thermodynamics of the
accelerating black holes in AdS spacetime was studied in Ref. \cite{The4}.
It was demonstrated that both the holographic computation and the method of
conformal completion yield the same result for the mass of these black
holes. The thermodynamic properties of charged rotating accelerating black
holes were evaluated in Refs. \cite{Anabalon2019,The7}. A new set of
chemical variables for the accelerating black hole was introduced in Ref. 
\cite{The5}. It was indicated how these expressions suggest that conical
defects emerging from a black hole can be considered as true hair. In
addition, it was discussed the impact of conical deficits on black hole
thermodynamics from this `chemical' perspective. So, the study of the
thermodynamic properties of accelerating AdS black holes in (non) extended
phase space has extracted a lot of interest (for example, see Refs. \cite%
{The2,The8a,The8b,The8c,The8}). Besides, accelerating black holes with a
conformally coupled scalar field in a magnetic universe have been studied in
Ref. \cite{Astorino2013}. Accelerating Kerr-Newman black holes were studied
in Ref. \cite{Astorino2016}. It was shown that, at extremality, it remained
a warped and twisted product of AdS. Besides, it was indicated that the
Kerr/CFT correspondence could be applied for the accelerating and magnetized
extremal black holes. Considering the $\mathbf{C}$-metric as a
gravitational field that describes an accelerating black hole in the
presence of a semi-infinite cosmic string (along the accelerating
direction), the gravitational energy enclosed by surfaces of constant radius
around the black hole was evaluated. The result revealed that the
gravitational energy of the semi-infinite cosmic string was negative. This
negative energy can explain the acceleration of the black hole, which moves
towards regions of lower gravitational energy along the string \cite%
{Carneiro2022}.

The modified theories of gravity can describe some of the phenomena that GR
could not explain them. One of these modified theories of gravity is
gravity's rainbow. To include Ultra-Violet (UV) completion of GR, gravity's
rainbow is introduced by Magueijo and Smolin \cite{Magueijo2004}. In
other words, in high-energy physics, and especially where the energy of
particles can be considered up to the Planck energy $E_{p}$, we need
alternative gravity to determine the effect of high energies on the
formulation of the system. The gravity's rainbow is a theory based on
quantum gravity. In the UV frequency (high-energy) limit, it causes
corrections in the energy-momentum dispersion relation \cite{Magueijo2004}.
The consequence of this modification is the dependence of spacetime on
energy, which recreates GR in the low frequency (IR) limit. Energy-dependent
rainbow functions in this gravity have the ability to influence some
properties of black holes and also compact objects. This theory of gravity
seems to be a promising candidate for dealing with UV divergences. In
gravity's rainbow, the test particles with various energy experience gravity
differently. Indeed, gravity has an effective behavior on particles
determined by their energy. In gravity's rainbow, there are two arbitrary
functions $f( \varepsilon)$ and $g( \varepsilon)$ (are called the rainbow
functions), which are added to spacetime. This theory of gravity introduced,
for example, the existence of remains of black holes after vaporization \cite%
{Ali2014}, a solution to the information paradox problem \cite%
{Ali2015b,Gim2015}, the existence of massive neutron stars \cite%
{NeutronS1,NeutronS2}, an explanation for the absence of black holes at the
LHC \cite{Ali2015}, the existence of a correspondence between
Horava-Lifshitz gravity and gravity's rainbow \cite{Garattini2015}. Besides,
in the context of black holes, various studies focusing on the effects of
rainbow functions on the thermodynamics of black holes have been made in
Refs. \cite{theGR1,theGR2,theGR3,theGR4,theGR5,theGR6,theGR7,theGR8,theGR9}.

Gravity's rainbow is a candidate for quantum gravity. On the other
hand, three-dimensional gravity provides a useful tool for studying quantum
gravity. According to the marvelous properties of accelerating black holes,
studying accelerating BTZ black holes in gravity's rainbow can give us
interesting information about the effects of high energy on the physics of
black holes, such as thermodynamic quantities (for example, temperature,
entropy, and heat capacity). It is notable that, in gravity's rainbow, the
effects of high energy appear in the metric. Since that we want to study
accelerating black holes, the effects of high energy appear in the $C$%
-metric (which is known as energy-dependent $C$-metric). It is noteworthy
that, the accelerating BTZ black holes in GR are introduced in some
literature \cite{Xu2012a,Xu2012b,EslamPanah2023,Astorino,Henriquez}. For
example, Astorino in Ref. GR \cite{Astorino}, first introduced accelerating
BTZ black holes, and then the thermodynamic properties of these black holes
were studied. In this regard, Xu et al. extended Astorino's work by adding a
topological constant to the accelerating BTZ black holes in GR \cite%
{Xu2012a,Xu2012b}. Besides, charged accelerating BTZ black holes in GR were
studied in Ref. \cite{EslamPanah2023}. Recently, Arenas-Henriquez et al.
constructed within three-dimensional GR a broad family of solutions
resembling the four-dimensional $C$-metric. Their analysis indicated that
the obtained geometries were much richer than the previous cases \cite%
{Henriquez}. Indeed, they found three classes of accelerating geometries and
interpreted their physical parameters by studying holographically.

In this paper, we want to extend accelerating BTZ black holes to
include energy-dependent spacetime. In other words, we are going to add the
effects of UV corrections to the accelerating BTZ black holes. For this
purpose, we will consider, the $C$-metric that introduced in Refs. \cite%
{Xu2012a,Xu2012b,Astorino}. Then, we will study the effects of rainbow
functions on the thermodynamic behavior of these black holes.

\section{Energy-dependent accelerating BTZ black holes}

Whereas we intend to extract the three-dimensional accelerating black hole
solutions in gravity's rainbow, so we first must create an energy-dependent $%
C-$metric.

To construct the energy-dependent metric, we follow \cite{Peng2008} 
\begin{equation}
h\left( \varepsilon \right) =\eta ^{\mu \nu }e_{\mu }\left( \varepsilon
\right) \otimes e_{\nu }\left( \varepsilon \right) ,
\end{equation}%
in which 
\begin{equation}
e_{0}\left( \varepsilon \right) =\frac{1}{f(\varepsilon )}\widetilde{e_{0}}%
,~~~\&~~~e_{i}\left( \varepsilon \right) =\frac{1}{g(\varepsilon )}%
\widetilde{e_{i}},  \label{apply}
\end{equation}%
where the tilde quantities refer to the energy-independent frame fields.
Considering the above conditions, we can create a suitable three-dimensional
energy-dependent $C$-metric for extracting accelerating BTZ black holes in
gravity's rainbow. Using the introduced $C$-metric\ in three-dimensional
spacetime \cite{Xu2012a,Xu2012b} and by applying Eq. (\ref{apply}), we can
extract a three-dimensional energy-dependent $C$-metric in the following
form 
\begin{equation}
ds^{2}=\frac{1}{\mathcal{K}^{2}\left( r,\theta \right) }\left[ -\frac{\psi
(r,\varepsilon )}{f^{2}(\varepsilon )}dt^{2}+\frac{1}{g^{2}(\varepsilon )}%
\left( \frac{dr^{2}}{\psi (r,\varepsilon )}+r^{2}d\theta ^{2}\right) \right]
,  \label{Metric}
\end{equation}%
where $\mathcal{K}\left( r,\theta \right) $ is the conformal factor in the
form $\mathcal{K}\left( r,\theta \right) =\alpha r\cosh \left( \sqrt{m-k}%
\theta \right) -1$. In addition, $k$ is related to topological constant
(which can be $\pm 1$\ or $0$). The coordinates of $t$ and $r$,
respectively, are in the ranges $-\infty <t<\infty $ and $0<r<\infty $. Due
to the lack of translational symmetry in $\theta $, we seem to have no
reason to restrict it to be in $\left[ -\pi ,+\pi \right] $. Therefore we
consider the range $-\pi \leq \theta \leq \pi $ (see Refs. \cite%
{Xu2012a,EslamPanah2023} for more details about the acceptable range of $%
\theta $). Notably, $\psi (r,\varepsilon )$\ is a radial energy-dependent
function that should be determined.

To answer whether the mentioned metric is under acceleration or not,
we can evaluate it by, i) a domain wall, which is studied in Ref. \cite%
{Henriquez}. ii) by calculating the proper acceleration, which is considered
in Refs. \cite{Xu2012a,Xu2012b,EslamPanah2023,Astorino}. To understand it,
we calculate the proper acceleration of the metric (\ref{Metric}).

We consider a static observer in this spacetime as $x^{\mu }\left(
\lambda \right) =\left( \lambda \frac{\mathcal{K}f(\varepsilon )}{\sqrt{\psi
(r,\varepsilon )}},r,\theta \right) $, where $\lambda $ is
the proper time. Notably, $\psi (r,\varepsilon )$ is the radial
energy-dependent function in which we are going to find it. By using the
definition of the proper velocity as $u^{\mu }=\frac{dx^{\mu }}{d\lambda }$, we get $u^{t}\left( \lambda \right) =\frac{\mathcal{K}%
f(\varepsilon )}{\sqrt{\psi (r,\varepsilon )}}$ (other components
of the proper velocity are zero, i.e., $u^{r}=u^{\theta }=0$). In
addition, The proper acceleration is defined as 
\begin{equation}
a^{\mu }=u^{\nu }\nabla _{\nu }u^{\mu },  \label{acceleration}
\end{equation}%
To find the proper acceleration, we need to $\psi (r,\varepsilon )$. So, after finding the radial energy-dependent function, we can get
the proper acceleration.

The three-dimensional action in gravity's rainbow with the cosmological
energy-dependent constant is given by 
\begin{equation}
\mathcal{I}(g_{\mu \nu })=\frac{1}{16\pi }\int_{\partial \mathcal{M}}d^{3}x%
\sqrt{-g\left( \varepsilon \right)}\left[ R\left( \varepsilon
\right)-2\Lambda\left( \varepsilon \right) \right] ,  \label{action}
\end{equation}%
where $R\left( \varepsilon \right)$ and $\Lambda\left( \varepsilon \right)$,
respectively, are Ricci scalar and the cosmological energy-dependent
constant, which depend on energy. We considered $G\left( \varepsilon
\right)=1$, in the above action.

Varying the action (\ref{action}) with respect to the gravitational field $%
g_{\mu \nu }\left( \varepsilon \right)$ yields 
\begin{equation}
G_{\mu \nu }\left( \varepsilon \right)+\Lambda\left( \varepsilon \right)
g_{\mu \nu }\left( \varepsilon \right)=0,  \label{Eqch1}
\end{equation}
where $G_{\mu \nu }\left( \varepsilon \right)$ is Einstein's tensor which
depends to energy.

Deformation of the standard energy-momentum relation as $E^{2}f^{2}(%
\varepsilon )-p^{2}g^{2}(\varepsilon )=m^{2}$ is one of the basics for
gravity's rainbow. In this relation, $\varepsilon =\frac{E}{E_{P}}$, and $%
E_{P}$\ is the Planck energy. Considering the mentioned upper limit for
energies that a particle can obtain (i.e., the energy of a particle cannot
be more than the Planck energy, $E\leq E_{P}$), we have $\varepsilon \leq 1$%
. In addition, $f\left( \varepsilon \right) $\ and $g\left( \varepsilon
\right) $\ are rainbow functions which satisfy the following conditions 
\begin{equation}
\underset{\varepsilon \rightarrow 0}{\lim }f\left( \varepsilon \right)
=1,~~~\&~~~\underset{\varepsilon \rightarrow 0}{\lim }g\left( \varepsilon
\right) =1,
\end{equation}%
where the above conditions ensure the standard energy-momentum relation in
the IR limit. Indeed, the deformation of the standard energy-momentum
relation turns to the standard case $E^{2}-p^{2}=m^{2}$, when $%
\varepsilon\rightarrow 0$. Notably, the aforementioned metric (Eq. (\ref%
{Metric})) reduces to three-dimensional $C$-metric in the IR limit, namely $%
ds^{2}=\frac{1}{\mathcal{K}^{2}\left( r,\theta \right) }\left[
-\psi(r)dt^{2}+\frac{dr^{2}}{\psi (r)}+r^{2}d\theta ^{2}\right] $, as we are
expected.

Considering the metric (\ref{Metric}), and the field equations (\ref{Eqch1}%
), one can get 
\begin{eqnarray}
eq_{tt} &=&eq_{rr}=\frac{\left( m-k+\psi -\frac{r}{2}\psi \prime \right)
\left( \mathcal{K}\left( r,\theta \right) +1\right) }{r^{2}\psi \mathcal{K}%
^{2}\left( r,\theta \right) }  \notag \\
&&+\frac{\psi \prime +2r\left( \alpha ^{2}\left( k-m\right) +\frac{\Lambda
\left( \varepsilon \right)}{g^{2}\left( \varepsilon \right) }\right) }{%
2r\psi \mathcal{K}^{2}\left( r,\theta \right) },  \label{qq1} \\
&&  \notag \\
eq_{\theta \theta } &=&\frac{\alpha ^{2}r^{2}\cosh ^{2}\left( \sqrt{m-k}%
\theta \right) \left( k-m-r\psi \prime +\frac{r^{2}\psi ^{\prime \prime }}{2}%
-\psi \right) }{\mathcal{K}^{2}\left( r,\theta \right) }  \notag \\
&&+\frac{\alpha r^{2}\cosh \left( \sqrt{m-k}\theta \right) \left( \psi
\prime -r\psi ^{\prime \prime }\right) }{\mathcal{K}^{2}\left( r,\theta
\right) }  \notag \\
&&+\frac{r^{2}\left( \frac{\psi ^{\prime \prime }}{2}+\alpha ^{2}\left(
k-m\right) +\frac{\Lambda \left( \varepsilon \right)}{g^{2}\left(
\varepsilon \right) }\right) }{\mathcal{K}^{2}\left( r,\theta \right) },
\label{eqq2}
\end{eqnarray}%
in which $\psi =\psi \left( r,\varepsilon \right) $. Also, the prime and
double prime are representing the first and second derivatives with respect
to $r$, respectively. In addition, $eq_{tt}$, $eq_{rr}$\ and $eq_{\theta
\theta }$, respectively,\ are related to components of $tt$, $rr$\ and $%
\theta \theta $\ of Eq. (\ref{Eqch1}). As one can see, $eq_{tt}$\ and $%
eq_{rr}$, are the same and they lead to Eq. (\ref{qq1}).

After some calculations, we can infer an exact solution of Eqs. (\ref{qq1})
and (\ref{eqq2}), which is given by 
\begin{eqnarray}
\psi \left( r,\varepsilon \right) &=&\mathcal{B}\alpha ^{2}\cosh ^{2}\left( 
\sqrt{m-k}\theta \right) r^{2}-\mathcal{B}_{1}r  \notag \\
&&+\frac{\frac{\Lambda \left( \varepsilon \right) }{g^{2}\left( \varepsilon
\right) }-\left( m-k\right) \alpha ^{2}}{\alpha ^{2}\cosh ^{2}\left( \sqrt{%
m-k}\theta \right) }+\mathcal{B}+k-m,  \label{f(r)Uch}
\end{eqnarray}%
where $\mathcal{B}$ is an integration constant, and 
\begin{equation}
\mathcal{B}_{1}=\frac{2\left( \frac{\Lambda \left( \varepsilon \right) }{%
g^{2}\left( \varepsilon \right) }-\left( m-k\right) \alpha ^{2}\right) +2%
\mathcal{B}\alpha ^{2}\cosh ^{2}\left( \sqrt{m-k}\theta \right) }{\alpha
\cosh \left( \sqrt{m-k}\theta \right) }.  \label{B1}
\end{equation}%
Notably, the obtained solution (\ref{f(r)Uch}), satisfies all components of
the field equation (\ref{Eqch1}).

To recover the obtained accelerating black holes in GR, we solve $\mathcal{B}_{1}$ versus $\mathcal{B}$ to omit the term $\mathcal{B}_{1}r$ in the solution (\ref{f(r)Uch}) because there is
no such term in accelerating BTZ black holes in GR. Solving Eq. (\ref{B1}),
we can obtain $\mathcal{B}$ in the following form 
\begin{equation}
\mathcal{B}=\frac{\left( m-k\right) \alpha ^{2}-\frac{\Lambda \left(
\varepsilon \right) }{g^{2}\left( \varepsilon \right) }}{\alpha ^{2}\cosh
^{2}\left( \sqrt{m-k}\theta \right) },  \label{C1}
\end{equation}%
and by replacing Eq. (\ref{C1}) within Eq. (\ref{f(r)Uch}), the
solution (\ref{f(r)Uch}) reduces to
\begin{equation}
\psi \left( r,\varepsilon \right) =\left( \left( m-k\right) \alpha ^{2}-%
\frac{\Lambda \left( \varepsilon \right) }{g^{2}\left( \varepsilon \right) }%
\right) r^{2}-m+k,  \label{f(r)C1}
\end{equation}%
in which cover the introduced solution in Refs. \cite%
{Xu2012a,Xu2012b,EslamPanah2023,Astorino}, when $g\left( \varepsilon \right)
=1$. In other words, in the IR limit, the aforementioned solution (\ref%
{f(r)C1}), turns into an accelerating BTZ black hole in GR \cite%
{Xu2012a,Xu2012b,EslamPanah2023,Astorino}. In addition, by considering $%
g\left( \varepsilon \right) =1$, and $\alpha =0$, the solution (\ref{f(r)C1}%
) reduces to the famous BTZ black hole provided $m>k$.

By considering the metric (\ref{Metric}), and by using Eq. (\ref%
{acceleration}) and the solution (\ref{f(r)C1}), we obtain the components of
the proper acceleration in the following forms
\begin{eqnarray}
a^{r} &=&\mathcal{K}\left( r,\theta \right) \left[ \left( k-m\right) \left(
\alpha r-\cosh \left( \sqrt{m-k}\theta \right) \right) g^{2}\left(
\varepsilon \right) \alpha \right]  \notag \\
&&+\Lambda \left( \varepsilon \right) r\mathcal{K}\left( r,\theta \right) ,
\label{ar} \\
&&  \notag \\
a^{\theta } &=&-\frac{\mathcal{K}\left( r,\theta \right) \sinh \left( \sqrt{%
m-k}\theta \right) \sqrt{m-k}g^{2}\left( \varepsilon \right) \alpha }{r},
\label{athe}
\end{eqnarray}%
and the $t-$component of $a^{\mu }$ is zero (i.e., $a^{t}=0$). So, the proper acceleration for the observer at $r=0$ is given  
\begin{equation}
\left\vert a\right\vert =\sqrt{a^{\mu }a_{\mu }}=\left( k-m\right) g\left(
\varepsilon \right) \alpha ,  \label{a}
\end{equation}%
where $\alpha $ can be related to the proper acceleration
at the origin when $m\neq k$. It is notable that, the magnitude $|a|$ of the proper acceleration depends on one of the rainbow functions (i.e., $g\left( \varepsilon \right) $), and it is due to the effects of gravity's rainbow (or high energy limit). Also, in the IR limit (i.e., $g\left( \varepsilon \right) =1$), $|a|$ turns to GR case, i.e., $\left\vert a\right\vert =\left( k-m\right) \alpha $, as we expected. According to the obtained component in the equations (\ref{ar}) and (\ref{athe}), $a^{\mu }$ is space-like, and we ought to have $a^{\mu }a_{\mu }>0$ in the static region of spacetime. Applying the equation (\ref{a}) for $m>k$, we see that the origin ($r=0$) is not in the static region. That the region of spacetime containing the origin $r=0$ is non-static is typical for black holes centered at the origin. For this reason, we infer it as an accelerating BTZ black hole which can exist only for $m>k$ (see Ref. \cite{Xu2012a}, for more details).

\textbf{Curvature Scalars:}\ Two well-known curvature invariants are the
Ricci and Kretschmann scalars. To get the singularity(ies) for the obtained
solutions in Eq. (\ref{f(r)C1}), we calculate the Ricci and Kretschmann
scalars which are given by 
\begin{eqnarray}
R &=&6\Lambda \left( \varepsilon \right) ,  \label{RiccUch} \\
&&  \notag \\
R_{\alpha \beta \gamma \delta }R^{\alpha \beta \gamma \delta } &=&12\Lambda
^{2}\left( \varepsilon \right) ,  \label{KreshUch1}
\end{eqnarray}%
where these quantities do not give any information about the singularity,
and are only related to the cosmological energy-dependent constant.

\textbf{Curvature Tensors:}\ The Ricci and Riemannian curvature tensors are
the most common ways to express the curvature of Riemannian manifolds. For
this purpose we obtain the components of $R^{\theta \theta }$ and $%
R^{t\theta t\theta }$ from three-dimensional energy-dependent $C$-metric
which are 
\begin{eqnarray}
R^{\theta \theta } &=&\frac{2\Lambda\left( \varepsilon \right) g^{2}\left(
\varepsilon \right) }{r^{2}}-\frac{4\Lambda\left( \varepsilon \right)
g^{2}\left( \varepsilon \right) \alpha \cosh \left( \sqrt{m-k}\theta \right) 
}{r}  \notag \\
&&+2\Lambda\left( \varepsilon \right) g^{2}\left( \varepsilon \right) \alpha
^{2}\cosh ^{2}\left( \sqrt{m-k}\theta \right) ,  \label{R11} \\
&&  \notag \\
R^{t\theta t\theta } &=&\frac{f^{2}\left( \varepsilon \right) g^{2}\left(
\varepsilon \right) \mathcal{K}^{4}\left( r,\theta \right) \Lambda\left(
\varepsilon \right) }{r^{4}\left( \left( m-k\right) \left( \frac{1}{r^{2}}%
-\alpha ^{2}\right) +\frac{\Lambda\left( \varepsilon \right) }{g^{2}\left(
\varepsilon \right) }\right) },  \label{RR11}
\end{eqnarray}%
where the aforementioned quantities diverge at $r=0$, i.e., 
\begin{eqnarray}
\underset{r\longrightarrow 0}{\lim }R^{\theta \theta } &\longrightarrow
&\infty , \\
&&  \notag \\
\underset{r\longrightarrow 0}{\lim }R^{t\theta t\theta } &\longrightarrow
&\infty ,
\end{eqnarray}%
so we encounter with a singularity at $r=0$.

Another important quantity that gives us information about black holes is
related to the event horizon. We find two roots for the metric function $%
\psi (r,\varepsilon )$ which are given by 
\begin{equation}
r_{root_{\pm }}=\pm \frac{\sqrt{\left( m-k\right) \left( \left( m-k\right)
\alpha ^{2}-\frac{\Lambda \left( \varepsilon \right) }{g^{2}\left(
\varepsilon \right) }\right) }}{\left( m-k\right) \alpha ^{2}-\frac{\Lambda
\left( \varepsilon \right) }{g^{2}\left( \varepsilon \right) }},
\label{rootuch}
\end{equation}%
where there is only a positive root which is $r_{root_{+}}=\frac{\sqrt{%
\left( m-k\right) \left( \left( m-k\right) \alpha ^{2}-\frac{\Lambda \left(
\varepsilon \right) }{g^{2}\left( \varepsilon \right) }\right) }}{\left(
m-k\right) \alpha ^{2}-\frac{\Lambda \left( \varepsilon \right) }{%
g^{2}\left( \varepsilon \right) }}$. In addition, to have a positive root we
must respect to 
\begin{equation}
\alpha ^{2}>\frac{\Lambda \left( \varepsilon \right) }{g^{2}\left(
\varepsilon \right) \left( m-k\right) },  \label{constraintI}
\end{equation}%
whereas $m>k$, the above constraint imposes two conditions on the
cosmological energy-dependent constant: i) This constraint (Eq. (\ref%
{constraintI})) always holds for the negative values of the cosmological
energy-dependent constant. ii) For the positive values of the cosmological
energy-dependent constant, we have to respect $\Lambda \left( \varepsilon
\right) <\alpha ^{2}g^{2}\left( \varepsilon \right) \left( m-k\right) $,
where indicates the existence of positive root is dependent on different
parameters of our system. In other words, the event horizon depends on the
acceleration parameter ($\alpha $), topological constant ($k$), the rainbow
function $g\left( \varepsilon \right) $, the cosmological energy-dependent
constant ($\Lambda \left( \varepsilon \right) $) and geometrical mass ($m$).

%%%%%%%%%%%%%%%%%%%%%%%%%%%%%%%%%%%%%%%%%%%%%%%%%%%%%%%%%%%%%%%
\begin{figure}[tbph]
\centering
\includegraphics[width=0.46\linewidth]{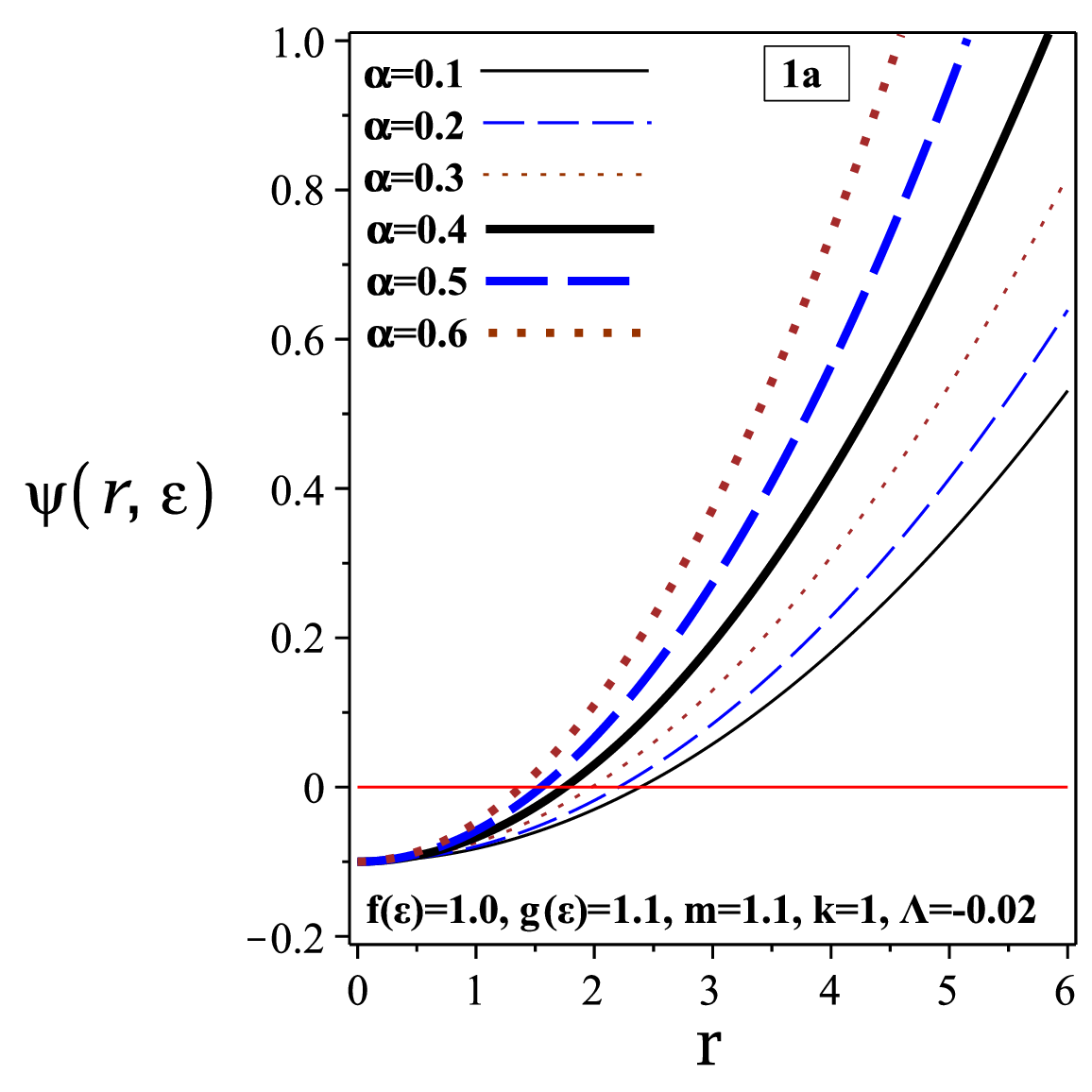} \includegraphics[width=0.46%
\linewidth]{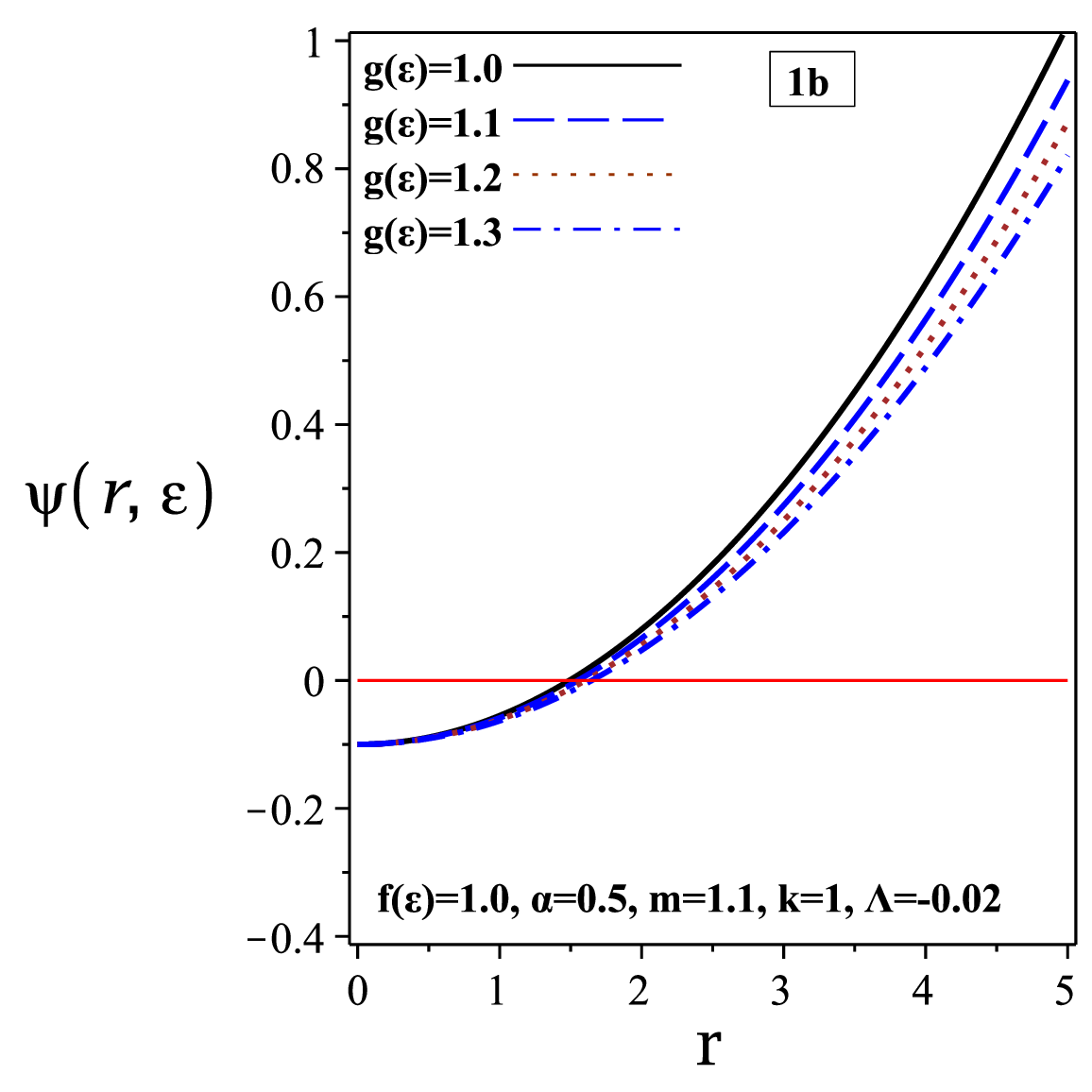} \newline
\includegraphics[width=0.46\linewidth]{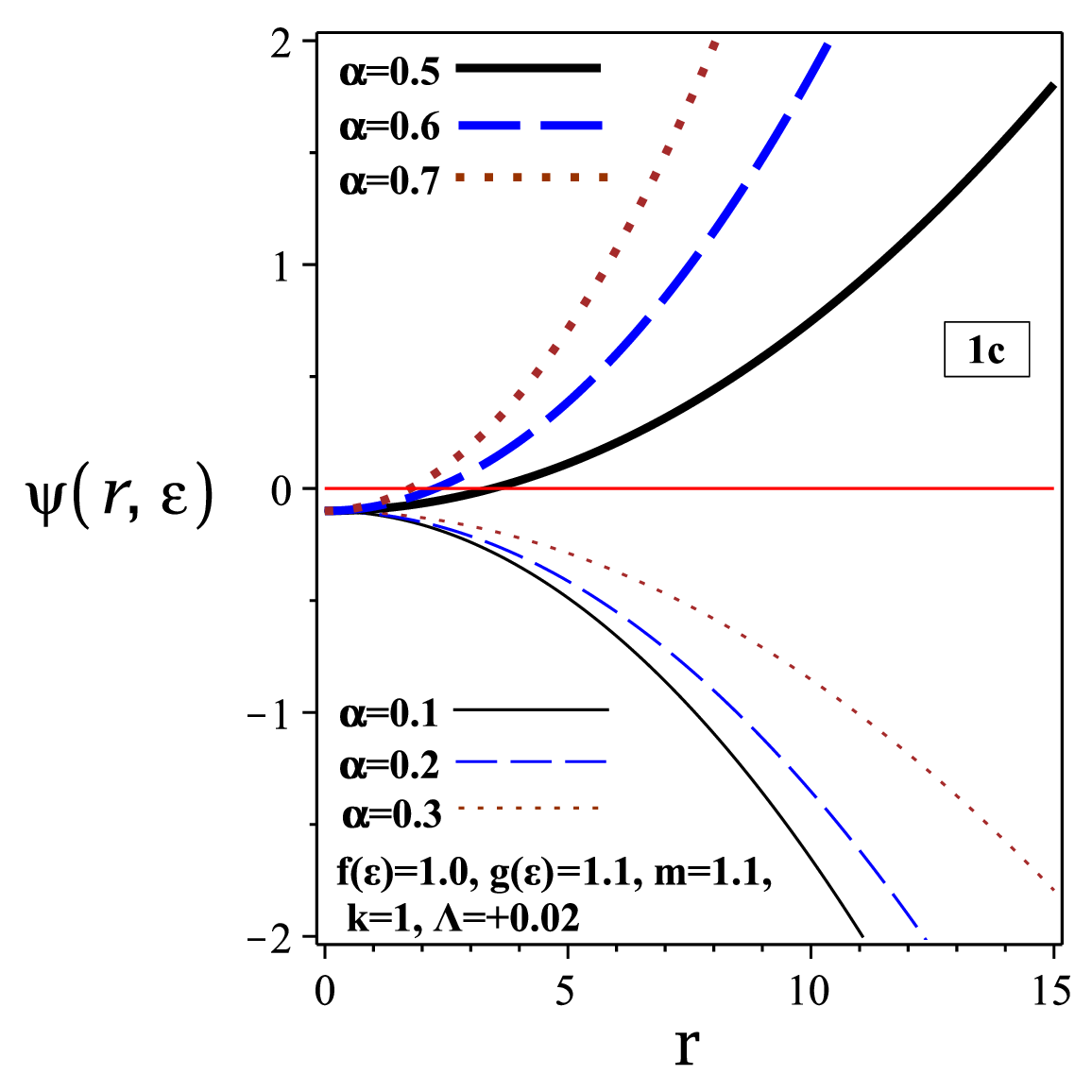} \includegraphics[width=0.46%
\linewidth]{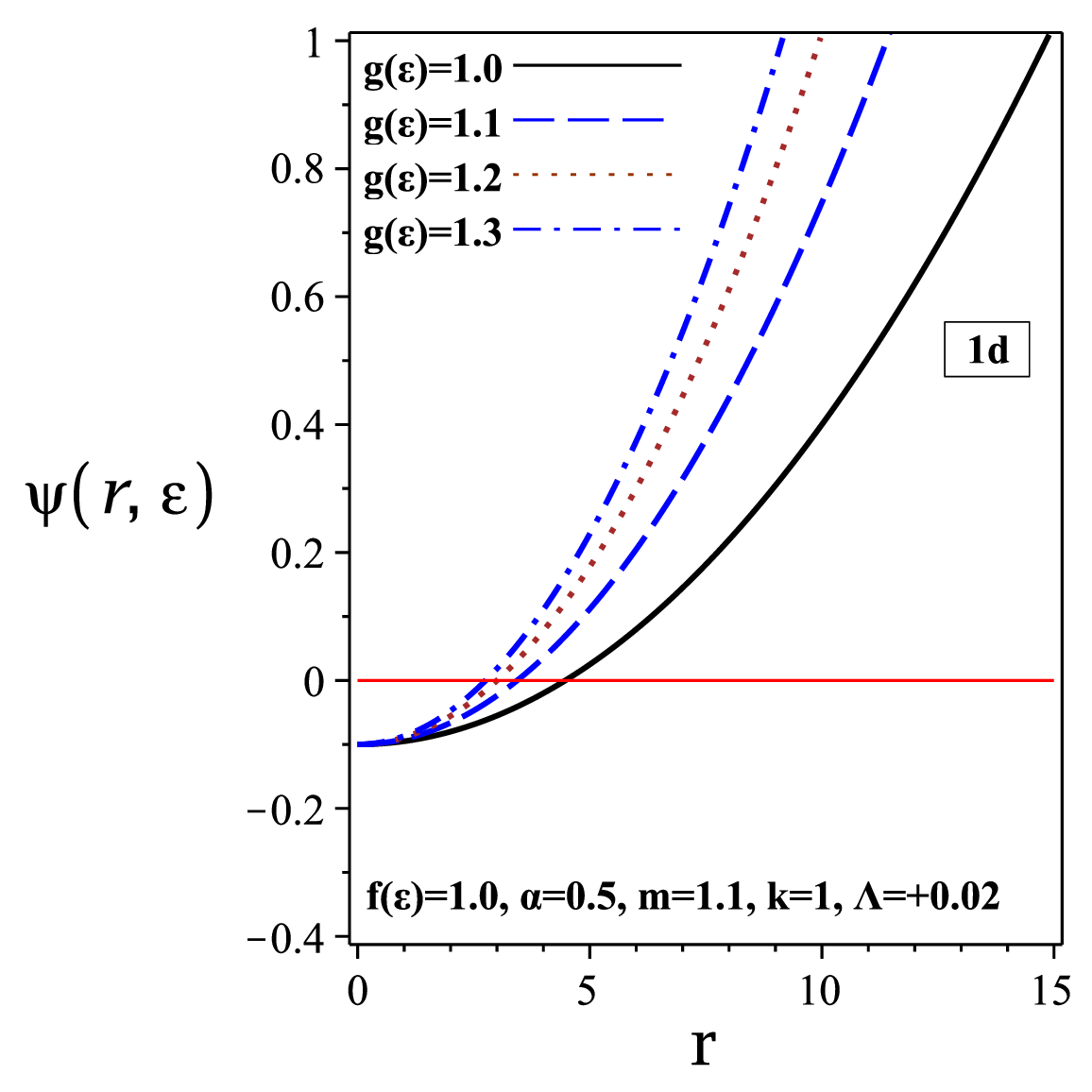} \newline
\caption{The metric function $\protect\psi (r,\protect\varepsilon)$ versus $%
r $ for different values of parameters. Up panels for the negative values of
cosmological energy-dependent constant. Down panels for the positive values
of the cosmological energy-dependent constant.}
\label{Fig1}
\end{figure}
%%%%%%%%%%%%%%%%%%%%%%%%%%%%%%%%%%%%%%%%%%%%%%%%%%%%%%%%%%%%%%%

Since there is a singularity at $r=0$, which may be covered by an event
horizon, the obtained solutions in Eq. (\ref{f(r)C1}) can be related to the
black hole solutions. Now, we are in a position to study the effects of
various parameters on the obtained black hole solutions. To get this
purpose, we plot the metric function (\ref{f(r)C1}) versus $r$\ in Fig. \ref%
{Fig1}. Here, we focus on the effects of the rainbow function ($%
g\left(\varepsilon \right) $) and the cosmological energy-dependent constant
on the behavior of black holes. Our findings include some interesting
behaviors for AdS black holes, which are

i) For $\Lambda\left( \varepsilon \right) <0$, and by increasing $\alpha $,
the event horizon decreases (see Fig. \ref{Fig1}a). In other words, large
black holes have small values of acceleration parameter.

ii) By increasing rainbow function, $g\left( \varepsilon \right) $, the
event horizon increases (see Fig. \ref{Fig1}b).

Indeed, slow accelerating black holes with large values of the rainbow
function are big black holes when $\Lambda\left( \varepsilon \right)<0$.

For accelerating dS black holes (or accelerating black holes with the
positive cosmological energy-dependent constant) we may encounter with two
kinds of different black holes, and it is related to the acceleration
parameter:

i) For $\alpha <\alpha _{critical}$, accelerating dS black holes do not have
any event horizon (naked singularity), see the three down diagrams in Fig. %
\ref{Fig1}c.

ii) For $\alpha >\alpha _{critical}$, accelerating dS black holes encounter
with an event horizon (see the three up diagrams in Fig. \ref{Fig1}c). In
this case, the event horizon decreases by increasing $\alpha $. In addition,
by increasing the rainbow function, the event horizon decreases (see Fig. %
\ref{Fig1}d).

In other words, there are two different behaviors which determine by $\alpha
_{critical}$. Slowly accelerating dS black holes have no event horizon, and
we encounter with naked singularity, but there is an event horizon for
accelerating dS black holes when $\alpha >\alpha _{critical}$.

\section{Temperature, Entropy and \textbf{Local Stability}}

Considering the black hole as a thermodynamic system, we are interested to
study the local stability for accelerating BTZ black holes in gravity's
rainbow. In other words, we evaluate the effects of rainbow functions, the
acceleration parameter, and the cosmological energy-dependent constant on
the local stability of such black holes. In order to calculate the local
stability, we have to obtain the heat capacity. For this purpose, we first
get Hawking temperature for these black holes. We obtain the geometrical mass ($m$) by equating $\psi (r,\varepsilon )=0$, in the
following form 
\begin{equation}
m=k+\frac{\Lambda \left( \varepsilon \right) r_{+}^{2}}{g^{2}\left(
\varepsilon \right) \left( \alpha ^{2}r_{+}^{2}-1\right) },  \label{m}
\end{equation}%
where $r_{+}$ is related to the event horizon of a black hole. To
extract the Hawking temperature, we employ the definition of surface gravity
as
\begin{equation}
T=\frac{\kappa }{2\pi },  \label{T}
\end{equation}%
where $\kappa =\sqrt{\frac{-1}{2}\left( \nabla _{\mu }\chi _{\nu
}\right) \left( \nabla ^{\mu }\chi ^{\nu }\right) }$, and $\chi
=\partial _{t}$ are the surface gravity and the Killing vector,
respectively. Here, we obtain the surface gravity, $\kappa $, by
using the metric (\ref{Metric}), which is given by
\begin{equation}
\kappa =\left. \frac{\sqrt{\mathcal{H}}g\left( \varepsilon \right) }{2%
\mathcal{K}\left( r,\theta \right) f\left( \varepsilon \right) }\right\vert
_{r=r_{+}},  \label{k}
\end{equation}%
in which $H$ is defined as
\begin{eqnarray}
\mathcal{H} &=&\left( r\psi ^{\prime }-2\psi \right) ^{2}\left( \mathcal{K}%
\left( r,\theta \right) +1\right) ^{2}+\psi ^{\prime 2}  \notag \\
&&  \notag \\
&&-2\psi ^{\prime }\left( r\psi ^{\prime }-2\psi \right) \left( \mathcal{K}%
\left( r,\theta \right) +1\right)  \notag \\
&&  \notag \\
&&-4\alpha ^{2}\left( k-m\right) \psi \sinh ^{2}\left( \sqrt{m-k}\theta
\right) .
\end{eqnarray}

Now, we are in a position to get the Hawking temperature. For this
purpose and by using the metric function (\ref{f(r)C1}) and Eqs. (\ref{m})-(%
\ref{k}), the Hawking temperature is
\begin{equation}
T=\frac{\Lambda \left( \varepsilon \right) r_{+}}{2\pi g\left( \varepsilon
\right) f\left( \varepsilon \right) \left( \alpha ^{2}r_{+}^{2}-1\right) },
\label{Tfinal}
\end{equation}%
where confirm that the Hawking temperature depends on all the parameters of
our system. In addition, there is a singularity for the temperature at $%
r_{+}=\frac{1}{\alpha }$. To remove this singularity, we omit this point.
Indeed, we do not permit to consider $r_{+}=\frac{1}{\alpha }$, because this
radius leads to a singularity in the temperature (\ref{Tfinal}).

According to the obtained temperature, we have two different behaviors for
AdS (or dS) black holes before and after the point $r_{+}=\frac{1}{\alpha }$%
. The temperature of accelerating AdS black holes with small radii (i.e., $%
r_{+}<\frac{1}{\alpha }$) are positive, but large black holes (i.e., $r_{+}>%
\frac{1}{\alpha }$) have the negative temperature. On the other hand, the
temperature of accelerating dS black holes with large radii (i.e., $r_{+}>%
\frac{1}{\alpha }$) are positive, but small black holes (i.e., $r_{+}<\frac{1%
}{\alpha }$) have the negative temperature. In addition, the existence of
rainbow functions does not affect this point. However, by decreasing these
functions, the positive area increases. As a result, the temperature can be
positive when $r_{+}>\frac{1}{\alpha }$ ($r_{+}<\frac{1}{\alpha }$) for
(A)dS black hole.

It was discussed in Refs. \cite{Xu2012a,EslamPanah2023}, the acceptable
range of $\theta $ may include the range $-\pi \leq \theta \leq \pi $, in
some certain cases. So, we can get the horizon area for these special cases
in the following form 
\begin{eqnarray}
\mathcal{A} &=&\left. \int_{-\pi }^{+\pi }\sqrt{g_{\theta \theta }}%
\right\vert _{r=r_{+}}  \notag \\
&=&\frac{-4\arctan \left( \frac{\left( \alpha r_{+}+1\right) \tanh \left( 
\frac{\pi \sqrt{m-k}}{2}\right) }{\sqrt{\alpha ^{2}r_{+}^{2}-1}}\right) r_{+}%
}{g\left( \varepsilon \right) \sqrt{\left( \alpha ^{2}r_{+}^{2}-1\right)
\left( m-k\right) }},  \label{AA}
\end{eqnarray}%
by replacing the horizon area (\ref{AA}) within $S=\frac{\mathcal{A}}{4}$,
we can get the entropy of accelerating BTZ black holes in gravity's rainbow
which is given by 
\begin{equation}
S=\frac{-\arctan \left( \frac{\left( \alpha r_{+}+1\right) \tanh \left( 
\frac{\pi \sqrt{m-k}}{2}\right) }{\sqrt{\alpha ^{2}r_{+}^{2}-1}}\right) r_{+}%
}{\pi g\left( \varepsilon \right) \sqrt{\left( \alpha ^{2}r_{+}^{2}-1\right)
\left( m-k\right) }}.  \label{S}
\end{equation}

Notably, in the absence of acceleration parameter (i.e., $\alpha =0$), the
aforementioned entropy reduces to $S=\frac{r_{+}}{2g\left(
\varepsilon\right) }$, as we expected. In addition, there is a singularity
for the entropy (\ref{S}) at $r_{+}=\frac{1}{\alpha }$, similar to the
Hawking temperature (Eq. (\ref{Tfinal})). It is clear that the singularity
depends on completely the acceleration parameter. As discussed in the
previous part, we do not consider the point $\frac{1}{\alpha }$.

According to the condition $m>k$, our analysis indicates that the entropy is
independent of the cosmological energy-dependent constant and is positive
when $r_{+}<\frac{1}{\alpha }$, and it is negative for $r_{+}>\frac{1}{%
\alpha }$. So, the accelerating BTZ black holes with small radii have
positive entropy.

Our findings of the temperature and entropy behaviors reveal that AdS BTZ
black holes can have positive temperature and entropy, simultaneously. As an
important result of this calculation, accelerating AdS BTZ black holes in
gravity's rainbow can be physical black holes.

In order to study the local stability of accelerating BTZ black holes in
gravity's rainbow, we calculate the heat capacity of such black holes. The
heat capacity is defined as 
\begin{equation}
C=\frac{T}{\left( \frac{\partial T}{\partial S}\right) }=\frac{T}{\left( 
\frac{\partial T}{\partial r_{+}}\right) /\left( \frac{\partial S}{\partial
r_{+}}\right) },  \label{Heat}
\end{equation}%
by considering the obtained temperature (\ref{Tfinal}) and the entropy (\ref%
{S}), and some calculations, we can get the heat capacity in the following
form 
\begin{equation}
C=\frac{\left( \mathcal{B}_{2}-\mathcal{B}_{3}\mathcal{B}_{4}\sqrt{\alpha
^{2}r_{+}^{2}-1}\right) r_{+}}{\pi g\left( \varepsilon \right) \sqrt{m-k}%
\left( \alpha ^{4}r_{+}^{4}-1\right) \mathcal{B}_{4}},  \label{C}
\end{equation}%
where $\mathcal{B}_{2}$, $\mathcal{B}_{3}$, and $\mathcal{B}_{4}$ are 
\begin{eqnarray}
\mathcal{B}_{2} &=&\left( \alpha ^{2}r_{+}^{2}-1\right) \tanh \left( \frac{%
\pi \sqrt{m-k}}{2}\right) \alpha r_{+},  \notag \\
\mathcal{B}_{3} &=&\arctan \left( \frac{\left( \alpha r_{+}+1\right) \tanh
\left( \frac{\pi \sqrt{m-k}}{2}\right) }{\sqrt{\alpha ^{2}r_{+}^{2}-1}}%
\right) , \\
\mathcal{B}_{4} &=&\left( \alpha r_{+}+1\right) \tanh ^{2}\left( \frac{\pi 
\sqrt{m-k}}{2}\right) +\alpha r_{+}-1.  \notag
\end{eqnarray}

\begin{figure}[tbph]
\centering
\includegraphics[width=0.46\linewidth]{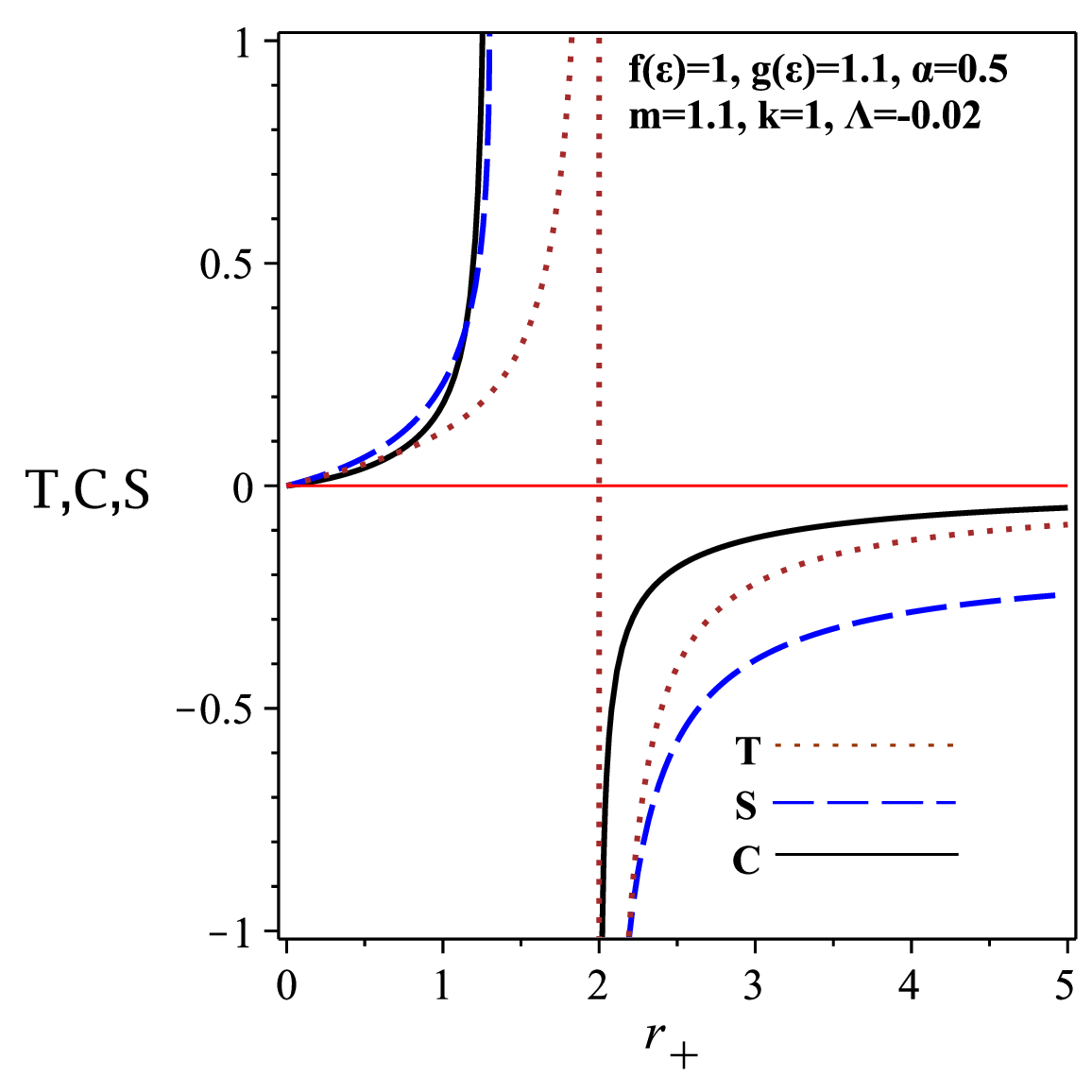} \includegraphics[width=0.46%
\linewidth]{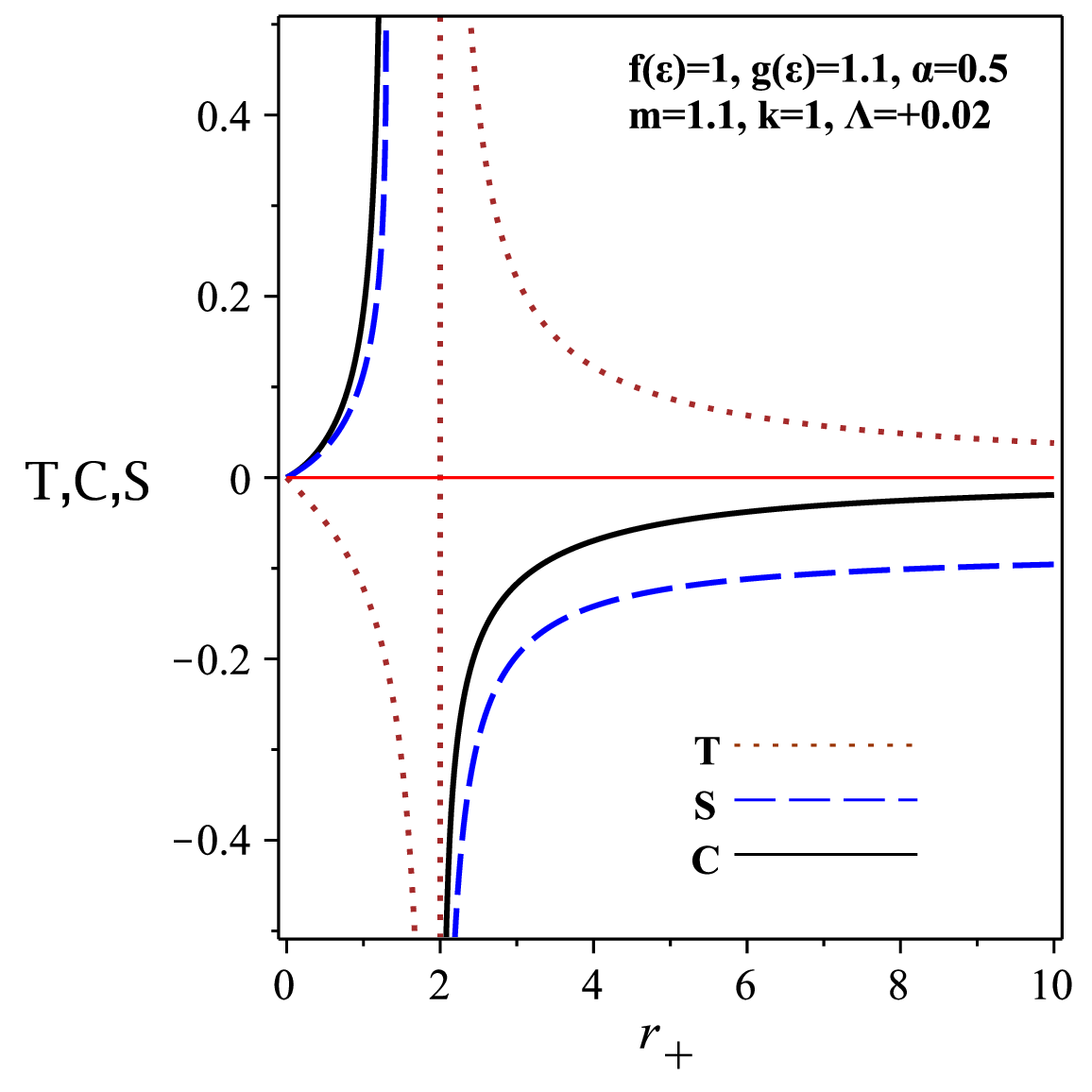} \newline
\caption{The temperature ($T$), entropy ($S$) and heat capacity ($C$) versus 
$r_{+}$ for AdS (left panel) and dS (right panel) cases, with different
values of parameters.}
\label{Fig2}
\end{figure}

To find the physical and local stability of the accelerating BTZ black holes
in gravity's rainbow, we plot the obtained heat capacity (\ref{C}) and the
temperature (\ref{Tfinal}) together versus the radius of black holes in Fig. %
\ref{Fig2}.

%%%%%%%%%%%%%%%%%%%%%%%%%%%%%%%%%%%%%%%%%%%%%%%%%%%%%%%%%%%%%%%
%%%%%%%%%%%%%%%%%%%%%%%%%%%%%%%%%%%%%%%%%%%%%%%%%%%%%%%%%%%%%%%

Our analysis includes two different areas which are related to the physical
and local stability of such black holes:

i) The temperature, entropy, and heat capacity of small accelerating AdS BTZ
black holes are always positive (see the left panel in Fig. \ref{Fig2}). So
the accelerating AdS BTZ black holes in gravity's rainbow are physical
objects and satisfy local stability ($C>0$) when $r_{+}<\frac{1}{\alpha }$.
But for $r_{+}>\frac{1}{\alpha }$, the accelerating AdS BTZ black holes
cannot be physical objects and also do not satisfy local stability. In other
words, accelerating AdS BTZ black holes with small radii are physical and
stable objects.

ii) For dS case, there is no area in which the positive temperature and the
positive heat capacity exist, simultaneously (see right panel in Fig. \ref%
{Fig2}). So, the accelerating dS BTZ black holes cannot be physical and
stable objects, simultaneously. The accelerating dS BTZ are physical objects
when $r_{+}<\frac{1}{\alpha }$, but the heat capacity is negative in this
area.

According to the obtained results of the temperature, entropy, and heat
capacity in the area $r_{+}<\frac{1}{\alpha }$, the accelerating BTZ black
holes in gravity's rainbow can be physical objects and satisfy the local
stability when the cosmological energy-dependent constant is negative. In
other words, accelerating AdS BTZ black holes with small radii are physical
objects and satisfy the local stability, simultaneously.

\section{\textbf{\noindent Summary and Conclusion}}

In this paper, we first constructed a three-dimensional energy-dependent $C$%
-metric. Then, by considering field equations, we found an exact solution.
Next, we evaluated this solution to find a singularity and roots. We found
that there was a singularity at $r=0$, which was covered by an event
horizon. These findings confined that the obtained solution was related to a
black hole solution. Then, we studied the behavior of accelerating BTZ black
holes under different parameters. There were two behaviors for accelerating
BTZ black holes, which were:

i) For $\Lambda\left( \varepsilon \right) <0$ (or AdS case):\ our analysis
indicated that BTZ black holes with small values of accelerating parameter
have large radii. In addition, the size of these black holes was sensitive
to rainbow function $g\left( \varepsilon \right) $. In other words,
accelerating AdS BTZ black holes had large radii when the effect of the
rainbow function increased.

ii) For $\Lambda\left( \varepsilon \right) >0$ (or dS case):\ there was a
critical value for acceleration parameter which less than this quantity
(i.e., $\alpha <\alpha _{critical}$), was not any root for BTZ black holes.
Indeed, we encounter with naked singularity when $\alpha <\alpha _{critical}$%
. For $\alpha>\alpha _{critical}$, accelerating dS BTZ black holes had one
root which decreases by increasing $\alpha $. Besides, by increasing the
rainbow function $g\left( \varepsilon \right) $, the radius of accelerating
dS BTZ black holes decreased.

We extracted the temperature of these black holes. Our findings indicated
that there were two different behaviors for both AdS and dS black holes,
before and after the point $r_{+}=\frac{1}{\alpha }$. In the range $r_{+}<%
\frac{1}{\alpha }$, the temperature of AdS black holes was positive, but it
was negative when $r_{+}>\frac{1}{\alpha }$. On the other hand, the
temperature of accelerating dS black holes was positive when $r_{+}>\frac{1}{%
\alpha }$, but it was negative when $r_{+}<\frac{1}{\alpha }$. Besides, the
rainbow functions appeared in the denominator of the temperature that
revealed the effects of gravity's rainbow on this quantity.

Then, we evaluated the entropy of these black holes. Our analysis indicated
that the entropy was positive when $r_{+}<\frac{1}{\alpha }$, and it was
negative for $r_{+}>\frac{1}{\alpha }$. So, the accelerating BTZ black holes
with small radii had positive entropy.

The last quantity that we calculated was related to the heat capacity. The
heat capacity was positive (negative) for $r_{+}<\frac{1}{\alpha }$ ( $r_{+}>%
\frac{1}{\alpha }$).

As a final result, our analysis of the temperature, entropy, and heat
capacity in the area $r_{+}<\frac{1}{\alpha }$, revealed that the
accelerating BTZ black holes in gravity's rainbow were physical objects and
satisfied the local stability when the cosmological energy-dependent
constant is negative. In other words, only small accelerating AdS BTZ black
holes were physical objects and satisfied the local stability,
simultaneously.

\begin{acknowledgements}
I thank an unknown referee for good comments and
advice that improved this paper. I would like to thank University of Mazandaran.
\end{acknowledgements}

\end{document}